# A Systematic Review of Knowledge Tracing and Large Language Models in Education: Opportunities, Issues, and Future Research

Yongwan Cho
Kalamazoo College
Kalamazoo, MI USA
Yongwan.cho20@kzoo.edu

Rabia Emhamed AlMamlook
Trine Univiersity
Angola, IN USA
Aalmamlookr@trine.edu

Tasnim Gharaibeh
Kalamazoo College
Kalamazoo, MI USA
Tasnim.Gharaibeh@kzoo.edu

Knowledge Tracing (KT) is a research field that aims to estimate a student's knowledge state through learning interactions—a crucial component of Intelligent Tutoring Systems (ITSs). Despite significant advancements, no current KT models excel in both predictive accuracy and interpretability. Meanwhile, Large Language Models (LLMs), pre-trained on vast natural language datasets, have emerged as powerful tools with immense potential in various educational applications. This systematic review explores the intersections, opportunities, and challenges of combining KT models and LLMs in educational contexts. The review first investigates LLM applications in education, including their adaptability to domain-specific content and ability to support personalized learning. It then examines the development and current state of KT models, from traditional to advanced approaches, aiming to uncover potential challenges that LLMs could mitigate. The core of this review focuses on integrating LLMs with KT, exploring three primary functions: addressing general concerns in KT fields, overcoming specific KT model limitations, and performing as KT models themselves. Our findings reveal that LLMs can be customized for specific educational tasks through tailor-making techniques such as in-context learning and agent-based approaches, effectively managing complex and unbalanced educational data. These models can enhance existing KT models' performance and solve cold-start problems by generating relevant features from question data. However, both current models depend heavily on structured, limited datasets, missing opportunities to use diverse educational data that could offer deeper insights into individual learners and support various educational settings.

## 1. INTRODUCTION

Advanced AI-related computing technologies—such as data analytics and deep learning—have enabled the collection and analysis of large-scale student data (Beverly, 2010; Ghosh et al., 2020). A key benefit of AI-enabled systems is their potential to significantly enhance learners' academic performance. Accurate modeling and prediction of knowledge, along with personalized guidance, can foster students' cognitive development and goal achievement while providing valuable insights for teachers, policymakers, and educational institutions (Huang et al., 2016; Divya et al., 2023). Recent advancements in pedagogical and computational methodologies have also led to diverse learning environments. These range from online learning platforms like Massive Open Online Courses and Khan Academy [1] to online question pools like LeetCode[2] and Codeforces[3]. Given these promising possibilities, various research communities—including educational institutions, local governments, and private companies—have shown considerable interest and invested heavily in AI-driven educational technology (Knox, 2020; Boninger et al., 2020).

Accordingly, intelligent tutoring systems, AI-supported learning platforms, are increasingly recognized for their ability to cater to diverse learning environments (Kabudi et al., 2021). Knowledge Tracing (KT), a fundamental task in developing these systems, aims to quantify a learner's knowledge level by tracking their responses to a previous sequence of questions and predicting their performance on new ones (Im et al., 2023). Traditional approaches to KT problems include Bayesian Knowledge Tracing (BKT) (Corbett and Anderson, 1995) and Factor Analysis Models (FAM) (Xiong et al., 2024). Recently, neural network architectures such as Recurrent Neural Networks (RNNs) (Lipton et al., 2015), Transformers (Vaswani et al., 2017), and Graph Neural Networks (Tong et al., 2020), have become common building blocks in knowledge tracing models (Choi et al., 2020). These models often incorporate auxiliary educational data like questions (Ghosh et al., 2020), knowledge concepts (Tong et al., 2020), and students' learning behaviors (Xu et al., 2023), to better model students' knowledge state and use them as inputs for KT models. Despite their remarkable performance, both traditional and deep learning-based models have limitations. Traditional models simplify exercise records, leading to information loss in the process of representing the exercises (Tong et al., 2020), and require manual annotation of questions and exercises, making implementation difficult (Li et al., 2024). Deep learning-based models, on the other hand, struggle with explainability, which results from the black-box nature inherent in neural network systems, and generalization to other teaching scenarios, such as open-ended questions (Li et al., 2024).

Meanwhile, the advent of Large Language Models (LLMs) has shown promising possibilities for educational applications. LLMs typically refer to transformer-based models with billions of parameters, featuring multi-head attention layers stacked in deep neural networks (Zhao et al., 2023). Their breakthroughs in education include prompting techniques for nuanced, context-aware responses (Shahriar et al., 2023; Brown et al., 2020; Wei et al., 2022), and advanced models like GPT-4—along with frameworks such as Divide-Conquer-Reasoning—to handle imbalanced educational datasets (Cui et al., 2024), often outperforming traditional Machine Learning (ML) techniques. Researchers have applied diverse methods to leverage LLMs in educational contexts, including few-shot learning (or context learning) (Li et al., 2024; Shahriar et al., 2023), further pre-training (Shen et al., 2021), and agent-based frameworks (Yang et al., 2024). These approaches aim to address LLMs' challenges, such as inefficient symbolic reasoning and hallucinations. However, the application of LLMs to Knowledge Tracing (KT) problems remains an emerging research field, with limited exploration to date (Li et al., 2024).

---

[1] https://www.khanacademy.org
[2] https://leetcode.com
[3] https://codeforces.com

Previous literature has extensively examined the technological advancements and pedagogical applications of AI in education, including intelligent tutoring systems (Ma et al., 2014). However, these studies often analyzed broad research trends without focusing on specific subjects like student performance. Notably, few review studies have explored the integration of various KT models across different settings, particularly in relation to LLMs. Despite the growing focus on LLMs, questions remain about refining their application in education. Many existing generative LLMs do not offer access to their model weights or APIs (Jung et al., 2024), and the technical specifics of fine-tuning methods are often unclear (Neshaei et al., 2024). This systematic review aims to synthesize the literature on the intersections, opportunities, and challenges of integrating KT models with LLMs in educational contexts. By offering an overview of various KT models and LLM techniques, and focusing on diverse educational contexts and student populations, this review can guide future research directions.

This paper investigates literature about knowledge tracing, large language models' application in education and the integration of KT and LLMs to address the following three research questions:

- RQ1. How can LLMs be applied in educational contexts, particularly in adapting to domain-specific contents in education and the capacity for supporting personalized learning?
- RQ2. What is the current state of KT models, from traditional to advanced approaches, and what potential challenges do they face that LLMs could address?
- RQ3. How can LLMs be integrated with KT models to: a) Address general concerns in the KT field? b) Overcome specific limitations of existing KT models? c) Function as KT models themselves?

In the following sections, we explore each of these research questions in depth. To address RQ1, we first investigate LLM applications in education, focusing on their adaptability to domain-specific contents in education and their capacity for supporting content generation and assistance, assessment and feedback, personalized learning, and educational tools and technologies. Next, we examine the development and current state of KT models—from traditional to advanced approaches—aiming to uncover potential challenges that LLMs could mitigate (RQ2). The core of this review centers on integrating LLMs with KT, exploring three primary functions: addressing general concerns in KT fields, overcoming specific KT model limitations, and performing as KT models themselves (RQ3).

## 2. SYSTEMATIC REVIEW METHODOLOGY

This review included only articles relevant to Knowledge Tracing and Large Language Models in educational contexts. The main search terms were "Knowledge Tracing," "Large Language Models,", "Education", and "Intelligent Tutoring System". While LLMs are a sub-branch of generative AI, this review specifically focused on LLMs to maintain a clear scope.

We searched for papers published from 2020 to 2024 using Google Scholar [15], arXiv [16], and ACM Digital Library [8], as well as other sources through manual search. The only exception not to restrict the publication dates is for RQ2, Knowledge Tracing models. This will allow for a broader historical context in understanding their development and persistent limitations of KT that LLMs could address. The effect measures primarily focused on KT model performance were accuracy, interpretability, and limitations. The measures for LLM integration in education focused on customization techniques and their applicability to addressing KT model limitations. The articles in this review addressed different target groups, such as K-12 and higher education. The review process adhered to the Preferred Reporting Items for Systematic Reviews and Meta-Analysis (PRISMA) guidelines, which are designed to guide systematic review and meta-analysis studies (Moher et al., 2009; Liberati et al., 2009).

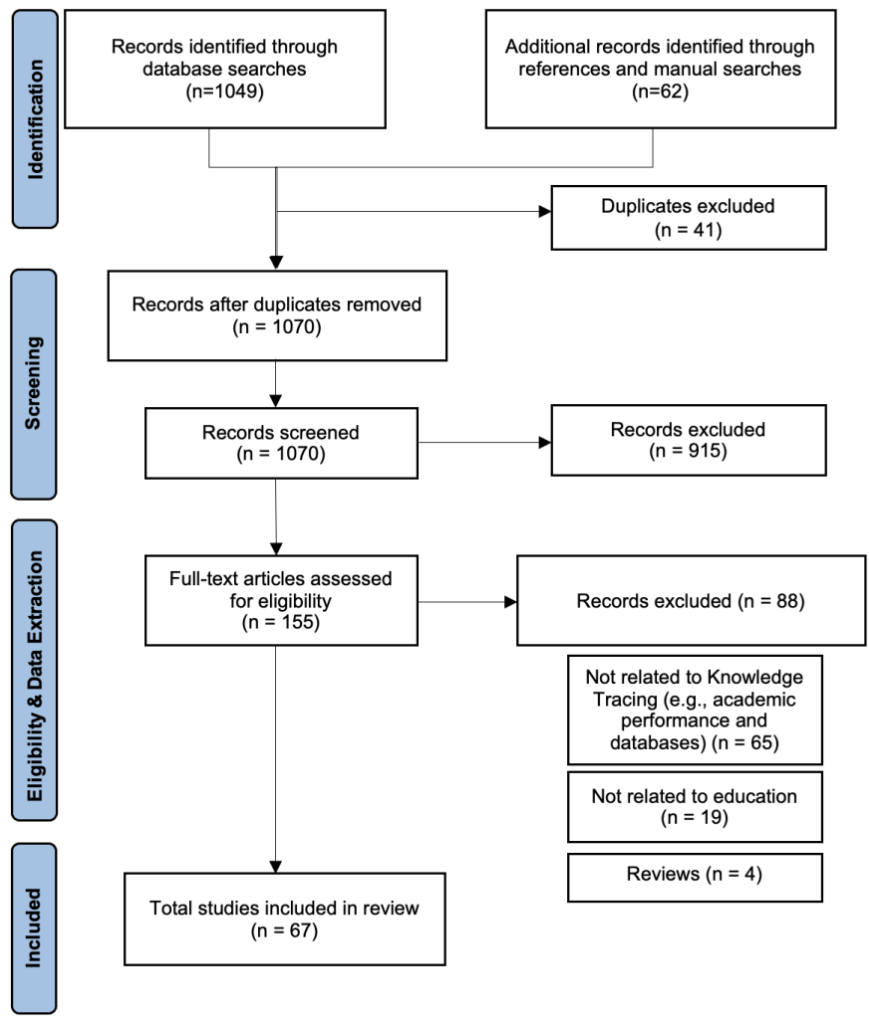

Figure 1. Flowchart of the articles' selection process

The two reviewers extracted data based on a predefined protocol that included outcomes of interest, study characteristics, and inclusion criteria. One reviewer conducted the primary review, while the other worked independently, holding weekly or bi-weekly meetings to discuss issues such as the inclusion and exclusion criteria. Screening and inclusion procedures were implemented to select articles for the primary analysis. Initially, 1049 articles were collected through the mentioned sources. Then, 41 duplicate articles were removed using manual searches. After applying filters in line with the search protocol shown in Figure 1, 67 studies fulfilled the inclusion criteria: 1. Studies related to student performances but not specifically to knowledge tracing, e.g., academic performances and databases. 2. Research not covering education. 3. Research written for review purposes. The full texts of the remaining articles were then reviewed. In the end, 67 articles were selected for analysis.

Table 1: Research questions and their motivations

| RQ# | Research questions | Motivations |
| --- | --- | --- |
| RQ1 | How can LLMs be applied in educational contexts, particularly in adapting to domain-specific contents in education and the capacity for supporting personalized learning? | Investigate the adaptability and application of LLMs in education. |
| RQ2 | What is the current state of KT models, from traditional to advanced approaches, and what potential challenges do they face that LLMs could address? | Investigate the current landscape of KT models, including their strengths and limitations. |
| RQ3 | How can LLMs be integrated with KT models to: a) Address general concerns in the KT field? b) Overcome specific limitations of existing KT models? c) Function as KT models themselves? | Delves into the integration of LLMs with KT models, aiming to enhance the overall functionality and effectiveness of KT systems |

## 3. RESULTS

### 3.1 RQ1 - LLMs application in education

The application of large language models in education can be classified into five main functions, including adaptability to domain-specific contexts in education, content generation and assistance, assessment and feedback, personalized learning support, and educational tools and technologies.

*Adaptability to domain-specific contexts in education*: Language models pre-trained on vast amounts of natural language data can be adapted for domain-specific or task-specific educational purposes. Transformer-based *models* like BERT and GPT excel in transfer learning, allowing them to adapt to new domains or tasks through further pre-training. For example, MathBERT, when further trained on a large mathematical corpus ranging from pre-kindergarten to mathematics paper abstracts, showed significant performance improvements over its base model on challenging educational tasks such as knowledge tracing and open-ended question scoring ([Shen et al., 2021](#)). As LLMs' capabilities have evolved, various techniques for utilizing them in the educational domain have also advanced. These include fine-tuning, which involves further training a pre-trained model on a specific dataset to adjust its parameters for a particular task or domain (McNichols et al., 2024); using distinct, or multi-agent, LLMs for role-playing simulations, such as students and professors(Phung et al., 2023; Yang et al., 2024); and few-shot learning, where the model is provided with a few examples of input-output pairs in the prompt to help it generalize and generate similar responses. These approaches have demonstrated powerful performance in tasks such as question-answering and arithmetic reasoning (Madaan and Yazdanbakhsh, 2022). However, models trained for domain or task-specific purposes often struggle if tasks or domains differ slightly from their trained purposes. For instance, MathBERT shows low performance in algebra error classification despite being further pre-trained on massive amounts of mathematics (Shen et al., 2021). Future research should aim to develop more generalizable techniques to train models across various educational domains.

*Content generation and assistance:* Researchers have explored diverse and advanced methods of employing LLMs to generate educational content. For instance, the approach described in (Shahriar et al., 2023) achieved significant improvements in quality and reduced hallucinations when generating middle school math explanations. They enhanced existing in-context learning (or few-shot learning) by focusing on structuring input contents(assertions-enhanced) and optimizing the number of examples. Additionally, LLMs have improved hint generation and validation. A classification model that selects more beneficial hints for K-12 science has leveraged LLMs to enhance performance by using labeled classification data as input (Zhang et al., 2024). (Phung et al., 2023) utilized a weaker LLM as a simulated student model for automatic hint validation of programming learning that is extracted from a stronger LLM. However, LLMs have shown limitations in certain areas of educational content generation, such as creating math multiple-choice questions with high-quality distractors that reflect common student misconceptions (Lee et al., 2024; Feng et al., 2024). Also, simple few-shot approaches have failed to generate high-quality explanatory content (Prihar et al., 2023). Future research should aim to expand LLMs' capabilities in content generation, particularly in areas where performance has been lacking, while also addressing the challenge of reproducibility by developing more accessible techniques for educators.

*Assessment and feedback*: LLMs show promise in reducing educators' workload by supporting assessments and providing student feedback. Pre-trained language models have made significant advances in automated scoring. For instance, fine-tuned BERTs have outperformed leading methods in math error classification with further pre-training (McNichols et al., 2023), open-response questions in science (Lee et al., 2023) and math (Baral et al., 2021), and have achieved human-level grading on reading comprehension (Fernandez et al., 2022). GPT-4, an advanced LLM, has demonstrated zero-shot capability to surpass existing Short Answer Grading (SAG) methods for elementary and college-level science education (Kortemeyer, 2024). Furthermore, high-quality and precise feedback can be generated for programming learning by using LLMs as validation mechanisms (Phung et al., 2023). Similarly, for middle school math, LLMs enhanced with data augmentation and preference optimization approaches can produce effective feedback (Phung et al., 2023). However, despite these advancements, LLMs still struggle with complex tasks such as generating feedback for open-ended math questions (McNichols et al., 2024) and Long Answer Grading(LAG) (Sonkar et al., 2024). These limitations indicate that LLMs are not yet capable of fully supporting sophisticated assessments. Future research should therefore focus on enhancing LLMs' ability to handle more complex, context-dependent educational tasks.

*Personalized learning support:* LLMs have been researched as personalized educational agents that provide learning experiences tailored to each student's knowledge state. These models aim to promote cognitive engagement and improve learning outcomes. For instance, (Abdelghani et al., 2023) demonstrated improved divergent question-asking performance among primary school students through prompt-based methods that generate linguistic and semantic cues. (Schmucker et al., 2024) implemented a learning-by-teaching format using two LLM-based agents: a student agent to whom learners explain biology topics, and a professor agent that provides support. Additionally, (Hou et al., 2024) addressed concerns about over-reliance on LLMs in programming education by providing a middle-stage product and implementing scaffolding methods—techniques that support students in learning skills beyond their current level. Furthermore, LLMs can enhance the efficiency of students' study plans by recommending the next concept to study based on their current knowledge state. In this application, LLMs provide deeper semantic meanings and relationships between concepts to the concept recommendation model, thereby improving its performance (Li et al., 2024). Despite the promising advancements, there is a need to improve supporting features such as assessment and content generation to enhance the quality of personalized learning experiences, as well as to develop an end-to-end model that integrates these features, enabling educators to effectively apply them in real-world classrooms.

*Educational Tools and Technologies:* The integration of LLMs addresses various concerns in both ITSs and conventional educational settings. LLMs have shown promise in tackling challenges posed by small, complex, and imbalanced educational datasets. For instance, (McClure et al., 2024) demonstrated that LLMs with assertions significantly outperformed ML models in accurately classifying cognitive engagement levels, indicating their potential in complex data analysis tasks. A major hurdle to widespread ITS adoption is the high cost of content development, which limits the diversity of topics covered. Knowledge Graphs (KGs)—a structured framework that organizes and connects educational content, concepts, and learner data through relationships—can help reduce costs by aligning models with educational domains. LLMs can further streamline this process by automating KG construction from unstructured text (Hatchett et al., 2024). Moreover, (Schmucker et al., 2024) developed an LLM-based Conversational Tutoring System (CTS) that reduces resource requirements for CTS content development. This system automates the generation of editable tutoring scripts from lesson texts, thereby enabling the application of CTS to a broader range of subjects. LLMs also show promise beyond the classroom. They can foster self-directed learning among students through chatbot applications (Ali et al., 2023). Additionally, LLMs are proving valuable in teacher professional development (PD). Through multi-agent LLM systems, they can assess teachers' content knowledge (CK) with human-like accuracy, addressing concerns over costs and limitations of time or location (Yang et al., 2024). However, future research must address challenges such as recency bias and the complexity of fine-tuning and prompt engineering for task adaptation, and the need for more accessible methods like prefix- (He et al., 2021) and adapter-tuning (Li and Liang, 2021). Furthermore, it is crucial to develop strategies that mitigate over-reliance on LLMs, ensuring their effective integration into educational settings where ML expertise and resources are limited.

### 3.2. RQ2 - Knowledge Tracing

#### 3.2.1 Traditional models

The two primary categories of traditional KT models are Bayesian Knowledge Tracing and Factor Analysis Models.

*Bayesian Knowledge Tracing:* BKT is a hidden Markov model with observable nodes that estimates a learner's knowledge states as binary variables across a set of skills, updating its state over time through Bayesian inference ([Bulut et al., 2023](#)). Each skill has four corresponding parameters: $P(L_0)$ for the Probability of concept mastery before learning, $P(T)$ for the Probability of transitioning from the not-known to the known state, $P(S)$ for the Probability of making a mistake (slipping) in a learned state, and $P(G)$ for the Probability of guessing an item correctly in an unlearned state. BKT then uses these parameters to estimate the probability of skill mastery. With expert labels associating questions and concepts, BKT can identify students' weaknesses and estimate future performance, enabling personalized intervention and support. Its use of Bayesian inference allows dynamic tracking of learning progress, whereas IRT, another traditional model, estimates students' ability at a specific time ([Corbett, 2001](#)). To address BKT's low estimation accuracy and limits to handle with complex datasets, researchers have explored various approaches, including integrating BKT with IRT ([Shanshan et al., 2017](#); [Pardos and Heffernan, 2011](#)), introducing student-specific parameters ([Pardos and Heffernan, 2010](#)), and utilizing skill hierarchies to handle questions involving multiple skills and inter-skill relationships ([Khajah et al., 2014](#)). Despite these efforts, BKT still faces challenges with performance on complex problem solving tasks and implementation due to its reliance on manual dataset annotations ([Li et al., 2024](#)).

*Factor Analysis Models:* FAM estimates students' future performance through logistic regression-based models that have proven their ability in this area by using various factors from students who solved problems ([Hakkal and Lahcen, 2024](#)). These models include Item Response Theory (IRT), which predicts student performance solely based on item difficulty ([Wilson and De Boeck, 2004](#)); Performance Factor Analysis (PFA), which uses counts of successful and unsuccessful attempts for each Knowledge Component (KC) without considering item difficulty, such that PFA can handle multi-skill questions and is as predictive as Deep Knowledge Tracing (DKT) ([Pavlik et al., 2009](#)); DAS3H, which combines IRT and DKT while capturing learning and forgetting of KCs ([Choffin et al., 2019](#)). A recent study used eXtreme Gradient Boosting (XGBoost), a scalable tree-boosting algorithm, to further enhance these logistic regression-based models ([Hakkal and Lahcen, 2024](#)). Notably, Knowledge Tracing Machine (KTM), which considers diverse factors than other FAMs—including students, items, skills, success attempts, and side information such as the learning environment—achieves comparable performance with Deep Learning KT models ([Vie and Kashima, 2018](#)). Similar to BKT, FAM requires manual dataset labels by experts ([Ghosh et al., 2020](#)), which provides interpretability beneficial in educational applications. However, this requirement also limits its implementation.

### 3.2.2 Deep Knowledge Tracing (DKT)

With the development of deep neural networks, researchers have taken several significant approaches for KT, including early DKT and memory-augmented models, attention-based models, and graph-based models.

*Early DKT and memory-augmented models:* Researchers have integrated deep neural networks to improve the performance of KT. ([Piech et al., 2015](#)) pioneered this approach by employing RNNs and Long Short-Term Memory (LSTM) ([Hochreiter and Schmidhuber, 1997](#)) to handle interaction sequences over time. Unlike BKT, DKT not only does not require manual skill-item mappings but also performs better without them, addressing an implementation issue of KT ([Shi et al., 2024](#)). However, early DKT models had limitations, including a lack of capability to capture complex Knowledge Components (KCs) learned by students ([Khajah et al., 2016](#)). To tackle the unrealistic assumption mentioned above, ([Sonkar et al., 2020](#)) exploited novel regularization to overcome overfitting and used initialization schemes inspired by the fastText architecture, instead of random embeddings that often show extremely poor performance in high dimensions. To capture complex KCs, ([Zhang et al., 2016](#)) suggested key-value memory networks as external memory structures, replacing the LSTM that summarizes a student's knowledge state in one hidden state. Furthermore, a modified LSTM for sequential modeling was employed while maintaining the same key-value memory structures ([Abdelrahman and Wang, 2019](#)). Despite the promising performance, the main issue with DKT is that its performance advantage comes at the cost of interpretability ([Shi et al., 2024](#)).

*Attention-based models:* Attention mechanisms, a cornerstone of the Transformer architecture ([Vaswani et al., 2017](#)), have recently been integrated into KT models ([Ghosh et al., 2020](#); [Piech et al., 2015](#); [Lipton et al., 2015](#); [Hochreiter and Schmidhuber, 1997](#); [Shi et al., 2024](#)), offering two primary advantages. Firstly, they enable the KT model to learn question-specific attention weights, reflecting each question's relative importance in predicting the probability of correct answers to subsequent questions. This addresses a key limitation of many traditional models, which assumes equal importance for all questions ([Minn et al., 2021](#)). Secondly, this approach enhances model interpretability by revealing the weights of specific inputs as main factors in predictions ([Pandey and Srivastava, 2020](#)), suggesting potential for personalized learning through automatic selection of appropriately difficult questions ([Ghosh et al., 2020](#)). Self-Attentive model for Knowledge Tracing(SAKT) ([Pandey and Karypis, 2019](#)) pioneered the use of attention mechanisms in KT. Attentive Knowledge Tracing(AKT) ([Ghosh et al., 2020](#)) calculates attention weights using time intervals between questions with an exponential decay rate, simulating the forgetting effect in learning. It also employs the Rasch model as concept-question embeddings to account for individual differences between questions, even within the same concept. Similarly, Relation-Aware Self-Attention for Knowledge Tracing(RKT) ([Pandey and Srivastava, 2020](#)) models forgetting behavior through an exponentially decaying kernel function while incorporating contextual information for self-attention. ([Im et al., 2023](#)) further emphasized forgetting behavior through linear bias, addressing the obscured influence of forgetting behavior from entanglements of question correlations. Separated Self-AttentIve Neural Knowledge Tracing(SAINT) ([Choi et al., 2020](#)) utilizes an encoder-decoder structure that separates exercises and responses in the input, allowing for distinct attention mechanisms tailored to each input type. While attention-based models outperform DKT and show promise for interpretability, there's room for improvement through exercise modeling—an area that remains under-explored ([Pandey and Srivastava, 2020](#)). Additionally, attention-based KT models still face challenges in generalizing to diverse real-world educational contexts, such as programming ([Shi et al., 2022](#)).

*Graph-based models:* Diverse relational structures exist between educational data, such as KCs, exercises, and students' knowledge states. Graph neural networks have recently gained attention for capturing these relational structures in KT problems ([Fan et al., 2014](#); [Seyler et al., 2017](#); [Phung et al., 2023](#)). Graph-based models can not only mine deeper inter-data relationships—a significant component for KT performance—but also learn probable influences between concepts and exercises, providing more interpretable and trackable conceptual graphs (Song et al., 2021). However, the relations between educational data can be defined in multiple ways (Tong et al., 2020), leading to the integration of various graph types in KT tasks. Hierarchical Exercise Graph for Knowledge Tracing(HGKT) (Tong et al., 2020) improves KT by leveraging hierarchical relations among exercises through a hierarchical exercise graph. This includes both direct and indirect support relations among exercises, providing a structured view of learning dependencies. Joint graph convolutional network based deep Knowledge Tracing(JKT) (Song et al., 2021) uses a joint graph convolutional network to incorporate multi-dimensional relationships. These include exercise-to-exercise and concept-to-concept connections, along with traditional exercise-to-concept relationships. This structure captures high-level semantic information, enhancing the model's prediction accuracy and interpretability, such as revealing underlying reasons for mastering certain knowledge. Session Graph-based Knowledge Tracing(SGKT) (Wu et al., 2022) incorporates session-based, dynamic graph structures that reflect students' real-time answering behaviors. It utilizes gated graph neural networks to model students' evolving knowledge states and Graph Convolutional Networks(GCN) to represent relations between exercises and skills. Despite these advancements, KT tasks still struggle with the cold start issue (Zhao et al., 2020) and require further research into interpretability and various educational contexts.

### 3.2.3 Recent approaches in KT problems

Recent works have utilized various ML models to address limitations in existing KT models. (Schmucker and Mitchell, 2022) explores transfer learning techniques to predict student performance in ITS when introducing new courses. This approach tackles the cold-start problem—where no initial student performance data is available for a new course—achieving prediction accuracy comparable to PFA that requires extensive student log data. Interpretable Knowledge Tracing(IKT) (Minn et al., 2021) employed a Tree Augmented Naive Bayes classifier, a simple extension of the Naive Bayes network, along with ML techniques to extract three latent features such as skill mastery, ability profile, and problem difficulty. Unlike deep learning-based models that use extensive parameters, IKT focused on these three features and managed better performance prediction while saving significant computational resources. However, some attempts leave room for improvement due to marginal performances. (Ghosh et al., 2021) explores extending KT methods beyond analyzing the correctness of student responses, focusing instead on predicting the exact options selected in multiple-choice questions. Extending several state-of-the-art methodologies, including LSTM network-based and attention network-based methods, yielded only marginal improvements, indicating a need for further research. Throughout the history of KT models, while some limitations such as implementation and interpretability have been somewhat addressed, generalizing KT models to other educational scenarios, such as open-ended KT and programming learning, remains a significant challenge.

### 3.3. RQ3 - Integration of LLM with KT

The three primary functions large language models can serve in knowledge tracing are: (i) addressing general concerns in the knowledge tracing fields, (ii) overcoming limitations of particular knowledge tracing models, and (iii) performing as knowledge tracing models.

*Addressing general concerns in Knowledge Tracing fields*: Large language models have recently been integrated with knowledge tracing models to improve performance by addressing limitations in the KT fields. For instance, LLMs can mitigate cold-start scenarios (Jung et al., 2024; Sonkar et al., 2024; Zhang et al., 2024), where performance prediction is hampered by limited data on students' problem-solving interactions, particularly when an intelligent tutoring system is in its early stages (Wu et al., 2022). Notably, KT models aligned with LLMs in cold-start scenarios not only demonstrated superior overall performance compared to traditional and deep learning-based models but also proved more generalizable across various domains, including mathematics, social studies, and science (Jung et al., 2024). Moreover, (Fu et al., 2024) achieved instant expansion of ITSs with newly ingested questions. This was accomplished by using LLMs to generate semantic and structural information on questions and concepts, enabling inductive knowledge tracing—a feat that existing ID-based KT models cannot achieve. However, extending these advancements to real-world teaching scenarios, such as open-ended questions and programming learning, remains a daunting task in KT fields, despite a few attempts (Li et al., 2024).

*Overcoming limits of particular Knowledge Tracing models*: Various existing KT models have leveraged LLMs to address their implementation challenges. For example, BKT, an ID-based model with greater interpretability than DKT, often struggles with limited data or requires manual high-quality annotations from pedagogical experts. Addressing this issue, (Li et al., 2024; Yang et al., 2022) showcased the effectiveness of LLMs' zero and few-shot capabilities in enabling knowledge tagging tasks, considering both explicit text semantic information and implicit relationships between knowledge concepts and question solutions. Furthermore, (Lee et al., 2023) demonstrated enhanced performance in state-of-the-art KT models such as DKT, Dynamic Key-Value Memory Networks for Knowledge Tracing(DKVMN), and AKT when combined with LLM-based difficulty predictions, surpassing their standalone counterparts. Moreover, (Liu et al., 2022) presents a novel approach to KT that focuses on open-ended responses, particularly in the context of computer science education. The students' knowledge estimation from existing KT models, like DKT and AKT, were utilized as inputs for GPT to generate predicted students' code submissions. Nevertheless, none of the assistance from the papers above addressed each KT models' more urgent and necessary concerns. Traditional KT models struggle with low performance, while deep neural network-based models need more applications beyond numerical predictions and face interpretability challenges due to their hidden vectors containing student knowledge (Li et al., 2024). Therefore, more research is needed to use LLMs to address each KT model's fundamental limitations and to improve diverse models, including attention-based ones that could benefit from better exercise modeling through LLMs, and KTM which offers relatively high performance and interpretability simultaneously.

*Performing as Knowledge Tracing models*: The advanced mathematical and logical inference capabilities of LLMs enable their extension to the domain of Knowledge Tracing. Fine-tuning relatively early LLMs, such as Generative Pre-trained Transformer(GPT)-2 and GPT-3, on extended prompts has achieved similar or higher performance than standard Bayesian models for Knowledge Tracing (Neshaei et al., 2024). Notably, advanced LLMs like GLM-4 and GPT-4 have not only demonstrated comparable or superior performance to state-of-the-art KT models such as DKT, AKT, and SAINT, but also provided natural language explanations (Li et al., 2024)—a significant achievement given that interpretability is a critical challenge in the field. However, educational data often contains manipulated information from students, presenting a unique challenge for LLMs in logical reasoning over falsified prompts. Moreover, the encoded data in parameters acts as a more dominant factor in the deduction process than the manipulated information in prompts (Yang et al., 2022). The research community lacks technical specifics of the exact fine-tuning methods provided by the models' owner companies (Neshaei et al., 2024). These studies indicate that LLMs' ability as KT models is at an experimental stage in technology and implementation, with a lack of appropriate training resources as the biggest challenge.

## 4. DISCUSSIONS

### 4.1. Summary of opportunities for LLMs' application in education and KT

The integration of LLMs in education offers promising advancements across multiple domains. This review highlights five primary functions of LLMs in education: adaptability to domain-specific contexts, content generation, assessment and feedback, personalized learning, and educational tools and technologies. Each function demonstrates LLMs' potential to enhance diverse learning environments and streamline processes that traditionally rely on human intervention.

#### 4.1.1. Five main educational domains of LLMs' applications

*Adaptability to Domain-Specific Contexts*: LLMs can be tailored for specific educational tasks through various techniques. These include further pre-training, fine-tuning, few-shot learning (in-context learning), and multi-agent-based frameworks. Models like MathBERT have shown improved performance in knowledge tracing and open-ended question scoring (Shen et al., 2021).

*Content Generation and Assistance*: Advanced methods in content creation, such as assertions-enhanced few-shot learning and the integration of weaker and stronger models for validation and generation, respectively, have demonstrated notable improvements in generating educational materials with reduced hallucination rates.

*Assessment and Feedback*: LLMs show promise in automating scoring and providing feedback across various subjects, achieving results close to human performance in specific domains. This is accomplished through techniques like enhancing LLMs with further pre-training and preference optimization, using LLMs as validation mechanisms, and even zero-shot methods that have surpassed SAG in certain areas.

*Personalized Learning Support*: LLMs have proven effective as virtual tutors, offering learning-by-teaching formats, adaptive study plans, and cognitive engagement tailored to students' knowledge states. They also address over-reliance concerns by providing middle-stage products with scaffolding methods.

*Educational Tools and Technologies*: LLMs can assist in handling resource-intensive tasks in ITS, analyzing imbalanced educational datasets, reducing barriers to broader application across disciplines, and providing valuable teacher professional development.

#### 4.1.2. Two major branches of KT models

Traditional KT models, such as BKT and FAM, play crucial roles in identifying and predicting future students' knowledge state. BKT uses Bayesian inference to track learning progression dynamically, while FAM—including IRT and PFA—primarily employs a logistic regression approach that considers educational data like item difficulty or both successful and unsuccessful student attempts. However, both BKT and FAM heavily rely on manual dataset labeling, limiting their scalability. Deep learning-based KT models, such as DKT, attention-based, and graph-based models, use neural networks to automate skill-item mappings and improve performance, but at the cost of interpretability. This trade-off suggests an area where LLMs could potentially bridge the gap between predictive power and interpretability.

*4.1.3. Three key ways to integrate KT and LLMs.*

LLMs can be integrated into KT models in three main ways: (1) *addressing general KT challenges*, (2) *overcoming specific KT model limitations*, and (3) *functioning as KT models themselves*. By tackling common KT issues like the cold-start problem, LLMs can boost performance across various subjects with minimal initial data. They can also enhance existing KT models such as BKT and DKT by generating difficulty predictions and knowledge tags, providing a more sophisticated understanding of students' knowledge states and reducing dependence on manually annotated data. Furthermore, LLMs can act as standalone KT models, offering logical inference and natural language explanations that deep neural networks lack. However, more research is needed to fine-tune their performance in complex educational settings.

*4.2 Challenges of integration between the two models and future research directions*

Despite the opportunities presented, several challenges persist:

*Limited research on LLMs in various KT models:* This review aims to explore the integration possibilities between KT models and LLMs. However, most studies have only utilized a few state-of-the-art models such as DKT, DKVMN, and AKT to evaluate LLMs' integration. The field is relatively new, with few studies attempting to integrate LLMs with KT models. Different KT models offer unique features and advantages, like interpretability and varying degrees of performance, and have distinct structures. Notably, while LLMs show promising potential to address KT model limitations, no studies have attempted to tackle each KT model's fundamental issues. Additionally, various question embedding methods, such as Rasch model-based embeddings (Ghosh et al., 2020), can significantly improve KT model performance. Further research exploring how LLMs might integrate with a broader range of KT models and embedding methods could provide a more comprehensive understanding of their role in advancing KT.

*Technical limitations:* Using LLMs for educational purposes comes with inherent technical challenges, especially when handling manipulated or imbalanced educational data that obscures student intent, particularly in real-world applications like open-ended questions and programming. For instance, LLMs suffer from "recency bias," overly relying on examples near the end of the prompt, thus biasing the output toward copying recent examples (Abdelghani et al., 2023). Addressing these challenges requires advancements in LLM interpretability and training methods tailored to educational data. However, the research community lacks technical specifics of the exact fine-tuning methods provided by model owners ([Neshaei et al., 2024](#)), with insufficient appropriate training resources being the biggest hurdle. Recently, OpenAI [4]launched fine-tuning for their model, GPT-4o, which is one of the most advanced model, on August 2024. Further research exploring this feature with domain-specific datasets can significantly improve performances at a lower cost.

*Cost and accessibility:* The computational expense of adapting pre-trained LLMs to specific downstream tasks limits their accessibility in educational contexts, as it primarily relies on fine-tuning. This process can be extremely costly and challenging for large models with an enormous number of parameters ([Raffel et al., 2019](#)). To address this, research should focus on cost-effective adaptation methods, such as parameter-efficient training techniques like prefix-tuning ([Li and Liang, 2021](#)) and adapter-tuning methods ([He et al., 2021](#)) that add only a few trainable parameters per new task, enabling broader application.

*Scalability and generalizability across educational contexts:* While effective in certain educational domains, LLMs face generalizability challenges when applied across diverse subject areas. Models often struggle with complex subject-specific tasks such as generating high-quality distractors for multiple-choice questions ([Lee et al., 2024](#); [Feng et al., 2024](#)), Long Answers Grading ([Sonkar et al., 2024](#)), and providing nuanced feedback in open-ended mathematics questions ([Moher et al., 2009](#)). This lack of generalizability constrains LLMs' applicability across varied educational contexts. To improve scalability and adaptability, researchers could integrate LLMs with conventional models for these tasks. Furthermore, creating shared educational datasets with diverse ages and subjects could facilitate the development of more universally applicable models.

*Improving educator accessibility and usability:* The technical expertise required to fine-tune or customize LLMs often creates a barrier to their widespread adoption in educational settings. Most researchers in the reviewed articles have strong engineering backgrounds and tend to focus on technological design and development. To address this, simplifying the integration process through user-friendly interfaces and customization options would empower educators to implement these models effectively. Moreover, providing comprehensive training and support on the pedagogical use of LLMs is crucial for fostering an environment where AI tools are perceived as accessible and beneficial teaching resources. Future research should explore innovative methodologies that actively involve teachers, students, and educational researchers.

*Lack of diverse educational databases and ethical concerns:* Current educational datasets are often limited in scope, with sparse data on individual student interactions, which challenges ID-based KT models that rely on extensive data points. For example, the popular MovieLens dataset in recommendation systems contains 25,000,095 ratings and 1,093,360 tag applications across 62,423 movies. In contrast, educational datasets like ASSIST094 have only 2,661 students with 165,455 interactions and 14,083 questions ([Fu et al., 2024](#)). Moreover, the lack of diverse educational datasets limits implementation for various educational environments, such as open-ended questions. Consequently, many reviewed studies had to create their own datasets. An effective KT model requires a more detailed set of student data, raising important privacy concerns. Further research is needed to determine what types of data should be used in AI models, carefully considering ethical issues (Sharma et al., 2019).

---

[4] https://openai.com/index/gpt-4o-fine-tuning/

## 5. CONCLUSION

Sixty-eight articles published between 2020 and 2024, examining the integration of KT and LLMs in education—with an extended timeline for KT model development (RQ2)—were reviewed. The findings reveal both promising opportunities and significant challenges in this emerging field of educational technology. While preliminary, the results provide a comprehensive overview of LLM integration in education across five main domains, the development and challenges of various KT models, and their integrations, considering various outcomes. The review highlights valuable trends and suggests potential research directions for researchers and practitioners. However, some limitations need to be considered. First, the studies reviewed were based on diverse datasets covering different educational levels, with some studies including multiple levels within the same dataset. This diversity introduces variability that may limit the generalizability of findings across educational contexts and levels. Second, the search strategy—focusing on specific term combinations like "Knowledge Tracing," "Large Language Models," and "Education"—may have excluded relevant articles using alternative terminology or emerging concepts in this rapidly evolving field.

## DECLARATION OF GENERATIVE AI SOFTWARE TOOLS IN THE WRITING PROCESS

*During the preparation of this work, the authors used Notion AI in all sections in order to identify grammatical errors and enhance clarity. After using this tool/service, the authors reviewed and edited the content as needed and take full responsibility for the content of the publication.*